\documentclass[printer, traditabstract]{aa}
\usepackage{setspace}
\usepackage{txfonts} 
\usepackage{textcomp}
\usepackage{natbib}
\usepackage{amssymb}
\usepackage{graphicx}
\usepackage{tabularx}
\usepackage[english]{babel}
\usepackage[T1]{fontenc}
\usepackage{multirow}

\def\HI{H{\,\small I}}
\newcommand{\sauron}{{\texttt {SAURON}}}
\newcommand{\atlas}{{ATLAS$^{\rm 3D}$}} 

\begin{document}
\title{The Lockman Hole project: gas and galaxy properties from a stacking experiment}

\author{K. Ger\'{e}b $^{1,2}$, R. Morganti$^{1,2}$, T.A. Oosterloo$^{1,2}$, G. Guglielmino$^{1,3,4}$, I. Prandoni$^{3}$}
\authorrunning{Ger\'{e}b et al.}

\titlerunning{\HI\ in Lockman Hole galaxies}

\institute{$^{1}$Netherlands Institute for Radio Astronomy (ASTRON), P.O. Box 2, 7990 AA Dwingeloo, The Netherlands \\ $^{2}$Kapteyn Astronomical Institute, University of Groningen, P.O. Box 800, 9700 AV Groningen, The Netherlands \\ $^{3}$Istituto di Radioastronomia, Bologna (INAF), via Gobetti 101, 40129 Bologna, Italy \\$^{4}$University of Bologna, Dept of Astronomy, Via Ranzani 2, Bologna, Italy}

\abstract{
We perform an \HI\ stacking analysis to study the relation between \HI\ content and optical/radio/IR properties of galaxies located in the Lockman Hole area. In the redshift range covered by the observations (up to $z = 0.09$), we use the SDSS to separate galaxies with different optical characteristics, and we exploit the deep L-band radio continuum image (with noise 11 $\mu$Jy beam$^{-1}$) to identify galaxies with radio continuum emission. Infrared  properties are extracted from the Spitzer catalog.\\
We detect \HI\ in blue galaxies, but \HI\ is also detected in the group of red galaxies  - albeit with smaller amounts than for the blue sample. We identify a group of optically inactive galaxies with early-type morphology that does not reveal any \HI\ and ionized gas. These inactive galaxies likely represent the genuine red and dead galaxies depleted of all gas. Unlike inactive galaxies, \HI\ is detected in red LINER-like objects. \\ 
Galaxies with radio continuum counterparts mostly belong to the sub-mJy population, whose objects are thought to be a mixture of star-forming galaxies and low-power AGNs. After using several AGN diagnostics, we conclude that the radio emission in the majority of our sub-mJy radio sources stems from star formation. \\
LINERs appear to separate into two groups based on IR properties and \HI\ content. LINERs with a 24 $\mu$m detection show relatively large amounts of \HI\ and are also often detected in radio continuum as a result of ongoing star formation. The LINER galaxies which are not detected at 24 $\mu$m are more like the optically inactive galaxies by being depleted of \HI\ gas and having no sign of star formation. Radio LINERs in the latter group are the best candidates for hosting low-luminosity radio AGN.}

\maketitle 

\section{Introduction}\label{Intro}

Cold gas is known to play an important role in the formation and evolution of galaxies \citep{Keres, Crain, Voort}. Therefore, our understanding of the structure, properties and evolution of galaxies  can not be complete without knowing about the various phases of the gas, including atomic hydrogen (\HI), in and around galaxies, and their dependence on other galaxy characteristics. The relation of \HI\ with luminosity, morphological type, color, environment, etc., has been widely investigated for the $z = 0$ Universe, using large single-dish \HI\ surveys, such as the \HI\ Parkes All Sky Survey (HIPASS) \citep{Meyer2004, Zwaan}, the Arecibo Legacy Fast ALFA (ALFALFA) survey \citep{Martin}, and detailed imaging surveys like WHISP \citep{Hulst}, THINGS \citep{THINGS}, \sauron\ \citep{Morganti2006, Oosterloo2010}, and \atlas\ \citep{Serra}.  

For work at moderate redshifts (i.e. $z$ $\sim$ 0.1), present-day radio telescopes are limited by low sensitivity, and directly detecting samples of galaxies requires huge investments of observing time \citep{Jaffe}.  A possibility of going beyond these sensitivity limitations and explore the gas content of galaxies at larger distances is given by stacking \HI\ profiles. Although limited so far, some studies using this technique have already been carried out \citep{Lah2007,Lah2009, Fabelloa, Fabellob, Verheijen, Delhaize}.  \\

Here we expand on these studies by using stacking techniques, to investigate the properties of the \HI\ in galaxies located in the Lockman Hole (LH) area, one of the well studied fields where multiwavelength data are available \citep{Fotopoulou, Gugli}, allowing different classes of objects to be identified and studied. In our study, broad-band radio observations aimed at obtaining deep continuum images are used to extract additional information about the \HI. This is the first time a {\sl combined analysis of line and deep continuum data} is attempted. A simultaneous line and continuum setup is becoming standard for current radio telescopes, and it will be even more common in future radio surveys. With this work we want to explore this possibility. Furthermore, the relatively high spatial resolution ($\sim 10^{\prime\prime}$) of the observations used in this study reduces the risk of confusion from companion galaxies, often present in previous, single-dish stacking experiments. 

The availability of deep radio continuum data (reaching 11 $\mu$Jy noise) allows us to investigate the relation between the gas (\HI) and the radio continuum emission in sub-mJy radio sources. The nature of the faint radio population responsible for the excess in number counts below $\sim 1$ mJy is controversial. Various studies (e.g. Simpson et al. 2006; Seymour et al. 2008; Smolcic et al. 2008, Mignano et al. 2008, Padovani et al. 2009) have identified this population with star-forming (SF) galaxies (starbursts, spirals or irregulars) and low power radio-loud and/or radio-quiet AGN (e.g. faint FR I, Seyfert galaxies), and many different approaches have been taken to disentangle these two phenomena (see Prandoni et al. 2009 for an overview). This study investigates the characteristics of sub-mJy sources and in particular their \HI\ content at low redshift, and it provides a first step in preparing the work to be done at higher redshift. 

Throughout this paper the standard cosmological model is used, with parameters $\Omega_m$ = 0.3, $\Lambda$ = 0.7 and $H_0$ = 70 km s$^{-1}$ Mpc$^{-1}$. 


\begin{figure}[t!]
    \begin{center}
      \includegraphics[width=0.45\textwidth]{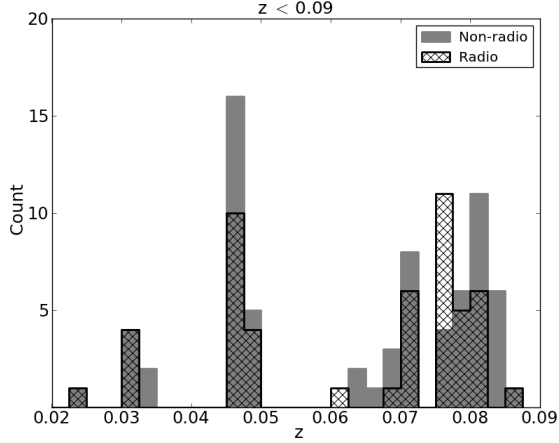}
   \end{center}
\caption{Redshift distribution of the SDSS galaxies in the Lockman Hole region, used for our \HI\ stacking experiment }
\label{fig:RedshiftHist}
\end{figure}

\section{The data: piggyback from the continuum observations}\label{SampleSelection}

For our \HI\ analysis we use data taken with the Westerbork Synthesis Radio Telescope (WSRT), originally aimed for the study of the radio continuum in the Lockman Hole area. 
The observations (Guglielmino et al. 2012, 2013 in prep.) were carried out at 1.4 GHz and centered on the coordinates R.A. = 10:52:16.6, Dec = +58:01:15 (J2000). An area of about  $6.6$ square degree  was covered by 16 pointings, each observed for 12 hour. Thanks to the deep observations, an rms noise of 11 $\mu$Jy beam$^{-1}$ was obtained in the central region of the final continuum mosaic, and about 6000 radio continuum sources were detected above a 5-$\sigma$ flux density threshold. The observations were optimized for the study of the radio continuum. However, as it is the case for many radio telescopes nowadays, the observations were carried out in spectral line mode, giving the possibility of deriving the \HI\ properties of the galaxies in the field to be. The setup uses $8 \times 20$ MHz bands (1300 - 1460 MHz) covering the redshift range 0 $\textless z \textless$ 0.09 with 512 frequency channels, corresponding to a velocity resolution of $\sim$75 km s$^{-1}$. The bands do not overlap, some gaps between bands are present to avoid well-known RFI-dominated regions.  

Relatively precise redshift measurements and sky positions are indispensable for stacking, thus we use the Sloan Digital Sky Survey (SDSS, York et al. 2000) catalog to select our spectroscopic galaxy sample. 

The $\sim$60 km s$^{-1}$ typical error in SDSS redshifts provides a suitable match with the Lockman Hole spectral resolution, making this dataset appropriate for stacking.
Up to $z = 0.09$, in total 120 SDSS galaxies can be used for stacking, 50 of them being associated with a radio source in the catalog of Guglielmino et al. (2012, 2013 in prep.). The distribution in redshift of the final sample is shown in Fig. \ref{fig:RedshiftHist}. 

In order to avoid biases and selection effects due to the relatively large redshift and luminosity distribution ($-23$ $<$ $M_r$ $<$ $-18$) of the objects, we limit our \HI\ analysis to the group of galaxies within 0.06 $<$ $z$ $<$ 0.09 (73 galaxies). This is also motivated by the larger number of objects (critical for stacking), and by the interest in terms of bridging future higher redshift studies. In the remainder of the paper we refer to the 0.06 $<$ $z$ $<$ 0.09 redshift range as the {\sl highest redshift bin}. 

\begin{figure}[t!]
    \begin{center}
        \includegraphics[width=0.42\textwidth]{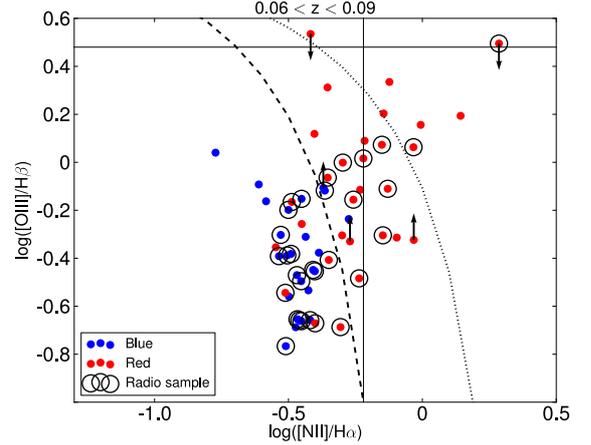}
   \end{center}
\caption{BPT diagram of 26 blue (SF+LINER) and 31 red (SF+LINER) Lockman Hole galaxies with SDSS spectra available in the highest redshift bin. Arrows mark the limits for galaxies for which only three optical lines are measured above S$/$N $>$ 2 (in the other galaxies all four lines have S/N $> 2$). The vertical and horizontal lines represent the conventional separation for LINERs (i.e. [\ion{O}{III}]/H$\beta$ $\textless$ 3 and [\ion{N}{II}]/H$\alpha$ $>$ 0.6) and Seyfert galaxies (i.e. [\ion{O}{III}]/H$\beta$ $>$ 3 and [\ion{N}{II}]/H$\alpha$ $>$ 0.6). In this work we adopt the demarcation by \cite{Kauffmann} ({\sl dashed line}) to separate star-forming galaxies from LINERs. The {\sl dotted line} indicates the more stringent demarcation to identify AGN, proposed by \cite{Kewley}. The radio subsample is marked by circles (31 sources). This figure excludes the inactive galaxies, see Sec \ref{sec:redLINER_inactive} for more detail.} 
\label{fig:BPT}
\end{figure}

\section{Characteristics of the galaxies in the redshift range 0.06 < z < 0.09}\label{SampleDescription}

In addition to the spectroscopic redshifts and positions, other galaxy parameters can be derived from the SDSS database. We use the SDSS DR8 Structured Query Language (SQL) tool\footnote{http://skyserver.sdss3.org/dr8/en/tools/search/sql.asp} to extract optical parameters for the Lockman Hole stacked sample. The {\sl g} and {\sl r} band magnitudes are extracted together with the [\ion{N}{II}], H${\alpha}$, [\ion{O}{III}], H$\beta$ line fluxes to constrain the galaxy colors and properties of the ionized gas.  
Blue and red galaxy samples are divided according to \cite{Blanton}, Lockman Hole galaxies with optical colors  $g - r$ $\textless$ $0.7$ are classified as blue (26 sources), while $g - r$  $>$ $0.7$ galaxies are classified as red (47 sources). Infrared (IR) information is extracted from the SWIRE Lockman Region 24 $\mu$m Spring '05 Spitzer Catalog\footnote{http://irsa.ipac.caltech.edu/cgi-bin/Gator/nph-dd?catalog=lockman\_24\_cat\_s05}. 

We separate galaxies with different optical line properties using the Baldwin, Phillips \& Terlevich (BPT) line ratio diagnostic diagram \citep{Baldwin}. According to the BPT diagnostics shown in Fig. \ref{fig:BPT}, our sample mainly consists of star-forming [\ion{H}{II}] galaxies and transition/LINER (Low Ionization Nuclear Emission Region) objects \citep{Ho1993, Heckman}. Star-forming galaxies are mostly associated with blue galaxies, while transition/LINER objects have red colors. Commonly, LINERs are defined to have [\ion{O}{III}]/H$\beta$ $\textless$ 3 and [\ion{N}{II}]/H$\alpha$ $>$ 0.6 (indicated by the vertical and horizontal solid lines in Fig. \ref{fig:BPT}). Various results from the literature indicate that LINER-type spectra, ionized by a harder continuum, can be obtained from different processes able to ionize the gas. Among these, the most often proposed mechanism is ionization by AGN, but more recent studies (Sarzi et al. 2010, and ref. therein), reveal that interaction of the warm ionized gas with the hot ISM and/or ionization by old pAGB stars is more likely. \cite{Sarzi} found that the role of AGN photoionization is confined to the central 2-3 arcsec, but only in a handful of galaxies. 

Given the fact that this class may include a variety of ionizing sources not necessarily dominated by an AGN, we generically define as {\sl LINERs} all galaxies above the \cite{Kauffmann} demarcation in Fig. \ref{fig:BPT} (dashed line). 

\begin{figure}[t!]
    \begin{center}
      \includegraphics[width=0.45\textwidth]{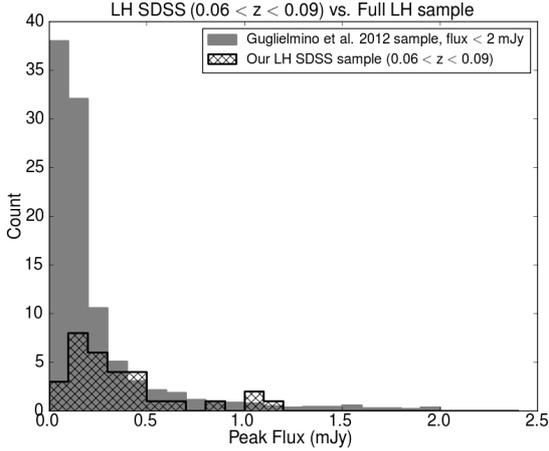}
        \end{center}
\caption{ Flux distribution of the highest redshift radio sample used in this study ({\sl black hatched region}, i.e. 31 radio sources with available SDSS redshift) compared to the distribution of the entire radio sample in the Lockman Hole ({\sl grey area}, Guglielmino et al. 2012). For the Guglielmino et al. (2012) sample, sources with larger fluxes are also found, however here we only plot sources with peak flux $<$ 2 mJy, in order to match the distribution of sources in this study. The latter distribution is normalized by the total number of galaxies (below 2 mJy) detected in the catalog of Guglielmino et al. 2012. The radio power corresponding to the fluxes is log(P$_{\rm 1.4GHz}$) $\textless$ 22.5 W Hz$^{-1}$. }
\label{fig:FluxDistr}
\end{figure}

Besides star-forming galaxies and LINERs, our red sample involves a third group of objects. We call these galaxies  {\sl optically inactive}, defined to have 
less than three optical lines ([\ion{N}{II}], H$\alpha$, [\ion{O}{III}], H$\beta$) detected with a S/N$ > 2$. For this reason, inactive galaxies do not appear in the BPT diagram. 

In addition to this, and particularly relevant in our analysis, in our SDSS-selected sample a group of objects is identified with a radio continuum counterpart (31 sources in the redshift range 0.06 $<$ $z$ $<$ 0.09). They are indicated with a circle in the BPT diagram of Fig. \ref{fig:BPT}. The 1.4 GHz radio flux distribution of these sources is shown in Fig. \ref{fig:FluxDistr} ({\sl black hatched region}). These objects mainly belong to the intriguing composite class of sub-mJy sources. At the distance of our galaxies ($z$ $<$ 0.09) these fluxes (S $<$ 2 mJy) correspond  to radio powers $\textless$ 10$^{22.5}$ W/Hz. By looking at the BPT diagram in Fig. \ref{fig:BPT}, we can already notice that the radio sources are spread among star-forming galaxies and LINERs. This further supports the composite nature of this faint radio population. 

Before proceeding, we investigate whether biases and selection effects are present in the SDSS-selected sample.  We start by showing in Fig. \ref{fig:FluxDistr} the distribution of radio fluxes in our selected sample compared to  those of the entire sample of radio sources in the Lockman Hole area. The two distributions are not completely identical: compared to the full sample, our sample contains only a small fraction of the faintest sources (S $<0.1$ mJy). Due to the radio continuum observations being deep, most of the missing sources in our SDSS-selected sample are probably higher redshift galaxies. This effect can also be due to the optical magnitude limit (r = 17.77) of the SDSS sample, together with the fact that the radio flux and optical magnitude tend to be correlated \citep{Condon, Mignano}.

\begin{figure}[t!]
    \begin{center}
      \includegraphics[width=0.45\textwidth]{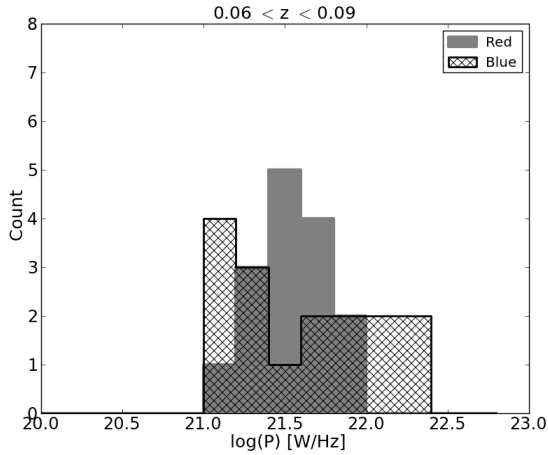}
   \end{center}
   \caption{Distribution of radio power for blue (16) and red (15) galaxies with radio counterparts in the highest redshift bin.}\label{fig:RadioPower}
\end{figure}

We also explore possible biases in optical luminosity and/or color as a function of redshift. The magnitude limit of the SDSS sample introduces a bias against low optical luminosity objects, and this bias is increasing with redshifts. However, the optical magnitude biases are not present in the highest redshift bin of our sample ($0.06$ $<$ $z$ $<$ $0.09$), where both blue and red galaxies (either radio-detected or not) show the same optical luminosity distribution in Fig. \ref{fig:Magn}. Red and blue galaxies with radio counterparts also show similar (even if not identical) radio power distributions (see Fig. \ref{fig:RadioPower}). 

Following the above described criteria, we define various samples of galaxies. Blue galaxies are classified as SF (23) and LINERs (3), while red galaxies can be classified into three groups: SF (7), LINERs (24) and inactive galaxies (16). {\sl We remark that the blue LINER and red SF samples are too small for stacking, therefore in this study we do not carry out any further analysis on these galaxies.}

In Sec. \ref{Sec:Radio} we use the Spitzer 3.6 $\mu m$, 4.5 $\mu m$, 5.8 $\mu m$ and 8.0 $\mu m$ fluxes to construct the IR color-color plot of the LH galaxies. In Sec. \ref{Sec:TwoGroupLINER} we also make use of the 24 $\mu m$ information. The sample of blue SF IR galaxies is complete to the 95 percent level (both in the four IR bands and at 24 $\mu m$). The sample of red LINERS is complete at 3.6 $\mu m$, 4.5 $\mu m$, 5.8 $\mu m$ and 8.0 $\mu m$, however the completeness drops to 60 percent at 24 $\mu m$. Almost 90 percent of the red inactive galaxies are detected in the four Spitzer bands, but none of the inactive galaxies has a 24 $\mu m$ detection.

\begin{figure}[t!]
    \begin{center}
      \includegraphics[width=0.45\textwidth]{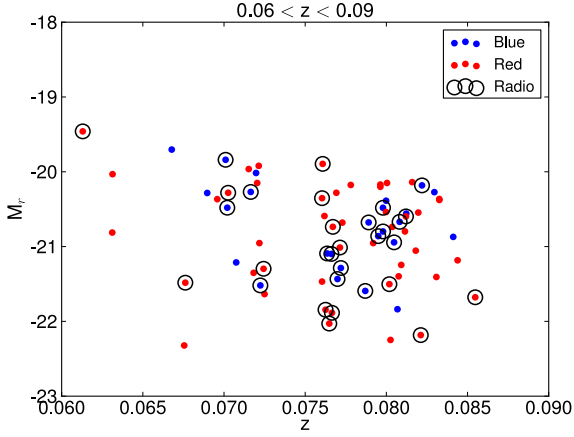}
   \end{center}
\caption{Magnitude of the blue (26) and red (47) galaxies in the highest redshift bin. Circles indicate radio detected objects.}\label{fig:Magn} 
\label{fig:Magn}
\end{figure}

\section{HI stacking}\label{stacking}

We used the radio-continuum calibrated {\sl uv}-data (from Guglielmino et al. 2013 in prep.) to create \HI\ data cubes of the 16 pointings. The cubes were produced after subtracting the continuum in the {\sl uv}-plane. The subtraction was achieved by using the clean components from the final continuum image.

The stacking script, written in {\sl Python}, performs a number of operations starting from the spectrum extraction in the data cubes at the location of the SDSS galaxies, a second continuum subtraction, de-redshifting, weighting of the spectra with the primary beam and noise, and making the final stack. 

We extract the spectra over an extended region around the optical position. One has to select this region carefully, for the reason that a small box may not include all of the \HI\ flux, while a larger region reaching beyond the \HI\ extent would only increase the noise and the risk of including companion galaxies.
After extensive tests, we derive a boxsize of 30 kpc and integrate the spectrum over a region corresponding to this radius for each galaxy. With larger boxsizes, the mass-luminosity ratios in the higher redshift bins do not change significantly, although the errors start to increase.

After the extraction of the spectrum, the stacking script performs a second continuum subtraction. This is needed, because after the first subtraction in the {\sl uv}-plane, continuum residuals remained in all cubes. The second subtraction is done directly in the extracted spectra by fitting a second order polynomial to the line-free channels, i.e. excluding a velocity range of 500 km s$^{-1}$  around the predicted position of the redshifted \HI\ frequency.  

To optimally weigh the stacked spectra, one has to apply corrections for the noise in every channel. This correction is needed, because the noise varies with frequency, becoming an important effect in our relatively large frequency range ($\sim$ 40 MHz, 0.06 $<$ $z$ $<$ 0.09).
For the Lockman Hole data, we create the cubes separately for every pointing, and handle them individually in our stacking procedure. We use the information on the noise along with the WSRT primary beam \citep{Popping} to correspondingly weigh the stacked spectra (Eq. \ref{eq:weighting}). The weighted sum of the source spectra, assuming a small range in redshift is:

\begin{equation}\label{eq:weighting}
 S( \nu ) = \frac{ \sum \frac{p_{i} (\nu)} {\sigma_{i}^{2} (\nu)} S_{i}(\nu)}  {\sum   (p_{i}^{2} (\nu) / \sigma_{i}^{2} (\nu) ) } 
\end{equation}

\noindent
where $S (\nu)$ is the stacked spectrum, $p_{i}(\nu)$, $\sigma_{i}(\nu)$ and $S_{i}(\nu)$ are the primary beam correction, the channel noise as a function of frequency and the extracted spectrum of source $i$. Flagged data and frequency gaps are given zero-weights. \\
In the stacking process, the noise of the co-added spectra is expected to decrease with $1/\sqrt{N}$, where $N$ is the number of stacked galaxies \citep{Fabelloa}. By stacking all 120 Lockman Hole objects with spectroscopic redshift we reach a noise level of $\sim$20 $\mu$Jy per channel, consistent with a factor of ten noise improvement compared to the initial $rms$ ($\sim$ 0.18 mJy/beam). 

From  the stacked profiles we derive the \HI\ mass. Considering a small redshift range, \HI\ masses can be derived from the formula: 

\begin{equation}
\frac{M_{HI}} {M_{\odot}} = \frac{235600}{(1+z)}  \bigg (\frac{S_{\nu}} {Jy}\bigg)    \bigg(\frac{d_{L}} {Mpc}\bigg)^{2}    \bigg (\frac{\Delta V}{km/s}\bigg)
\end{equation}\label{eq:HImass}

\noindent
where $z$ is the mean redshift of the stacked sample, $S_{\nu}$ is the average flux integrated over the $\Delta V$  velocity width in the emitter's frame of the \HI\ profile and $d_{L}$ is the average luminosity distance. \HI\ masses are further used for mass-luminosity evaluation, where the luminosities are calculated from SDSS $r$ band magnitudes.

\section{Results}\label{results}
In Table \ref{SF_LINER_HI_Table} we list the values of \HI\ mass and M$_{HI}$/L$_r$ for all the groups we examine. The related results are discussed in the following sections.
 

\begin{table*}[!]
   \begin{center}
   
     \begin{tabular}{c c c c c c}
                           &         Section	 \ref{Sec:ColorSep}						  &       \\
          \hline
                                                          & $M_{\rm HI}$ $(10^{9}$ $\rm M_{\odot})$    & $M{_{\rm HI}}/L_r$ ($\rm M_{\odot}  / \rm L_{\odot} )$ \\
          \hline                            
        
                Blue (SF+LINER) (26)      &    6.12 $\pm$ 0.40    &   0.38 $\pm$ 0.02  \\
                Red (SF+LINER+Inactive) (47)     &    1.80 $\pm$ 0.20   &   0.08 $\pm$ 0.01 \\
           \\     
                     &         Section	\ref{sec:redLINER_inactive}						  &       \\
     \hline
                                                             & $M_{\rm HI}$ $(10^{9}$ $\rm M_{\odot})$    & $M{_{\rm HI}}/L_r$ ($\rm M_{\odot}  / \rm L_{\odot} )$ \\
                
  \hline

                 Red LINER (24)    &    2.40 $\pm$ 0.45     &   0.09 $\pm$ 0.02  \\
                 Red Inactive (16)  &  $<$  0.95 $\pm$ 0.32     & $<$  0.06 $\pm$ 0.02 \\
        \\                                       
                 &         Section \ref{Sec:TwoGroupLINER}						  &       \\                    
  \hline
                                                                 & $M_{\rm HI}$ $(10^{9}$ $\rm M_{\odot})$    & $M{_{\rm HI}}/L_r$ ($\rm M_{\odot}  / \rm L_{\odot} )$ \\
 
  \hline
                  LINER 'IR SF ' (14)         &       3.62 $\pm$ 0.94      &   0.14 $\pm$ 0.03  \\
                  LINER 'IR inactive' (10)   &  $<$  1.20 $\pm$ 0.40  & $<$ 0.04  $\pm$ 0.015 \\
  
  \end{tabular}
    \caption{\HI\ content and 3-$\sigma$ upper limits derived for various groups in the highest redshift bin. The number of stacked sources is indicated in the brackets in Col. 1. Related results are discussed in the denoted sections.}\label{SF_LINER_HI_Table} 
\end{center}
\end{table*} 			

\subsection{HI content and color separation}\label{Sec:ColorSep}

\begin{figure}[t!]
    \begin{center}
      \includegraphics[width=0.45\textwidth]{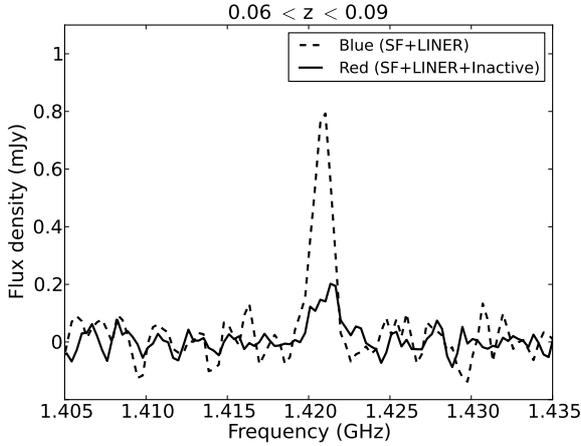}
   \end{center}
   \caption{\HI\ spectra obtained for blue ({\sl dashed line}) and red galaxies ({\sl solid line})} 
\label{fig:RedBlue}
\end{figure}

Blue galaxies (26 objects) are almost all star-forming (23 SF) as can be seen in Fig. \ref{fig:BPT}. As presented in Fig. \ref{fig:RedBlue} and Table \ref{SF_LINER_HI_Table}, the blue sample is detected with the highest \HI\ content.

For red galaxies (47 in total) we find lower \HI\ masses and mass-luminosity ratios (see Fig. \ref{fig:RedBlue} and Table \ref{SF_LINER_HI_Table}). We compare the numbers detected at $z$ = 0.09 for the red sample, with the \HI\ content derived by \cite{Serra} for early-type galaxies within 42 Mpc distance (corresponding to redshift z = 0.009). These nearby early-type galaxies are found to span a large range of \HI\ mass-luminosity, making it difficult to do a direct comparison with values from a stacking experiment as the one presented here. 
Because galaxies in our sample tend to be optically brighter compared to \atlas, we apply a luminosity cut (4.5 $\times$ 10$^{9}$ L$_\odot$) for \atlas\ galaxies in order to match the LH sample. 

The average mass-luminosity $M/L_r$  = 0.003 ($\rm M_{\odot}  / \rm L_{\odot} )$ of \atlas\ galaxies is lower compared to the red LH sample (Table \ref{SF_LINER_HI_Table}).  The difference in \HI\ mass-luminosity ratios likely suggests that, compared to \atlas, our color-selected LH sample may still contain a variety of morphological types, including more \HI -rich late-type galaxies. The possibility will be further investigated in Sec. \ref{Sec:TwoGroupLINER}.

The above presented red sample includes all the red galaxies (according to the classification presented in Sec. \ref{SampleDescription}). In the next section we investigate in more detail the \HI\ properties of the two main red groups, namely LINERs and inactive galaxies.

\subsection{The connection between cold and ionized gas in red galaxies}\label{sec:redLINER_inactive}

As can be seen in Table \ref{SF_LINER_HI_Table}, the \HI\ mass of red LINERs is higher compared to the entire sample of red galaxies, but the \HI\ mass-luminosity ratios are similar.

\begin{figure}[t!]
    \begin{center}
      \includegraphics[width=0.45\textwidth]{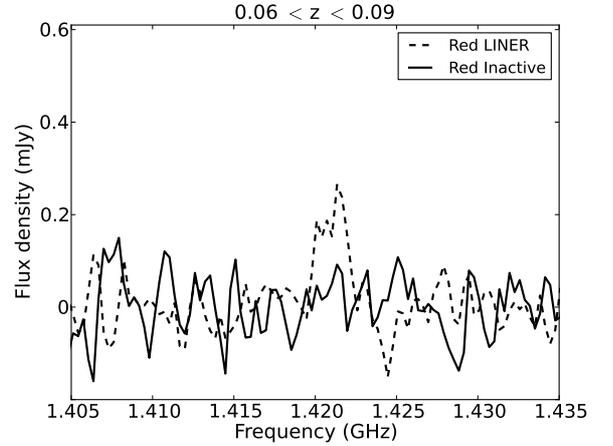}
   \end{center}
     \caption{Stacked spectra of red LINERs ({\sl dashed line}) and red inactive galaxies ({\sl solid line})}
\label{fig:Profiles2}
\end{figure}		

An intriguing result of this stacking exercise is that while red LINER galaxies (24 objects) are clearly detected in the stacked profile, the group of {\sl inactive} galaxies does not appear to show an \HI\ gas detection (see Fig. \ref{fig:Profiles2} and Table \ref{SF_LINER_HI_Table}). Stacking of this group (16 objects) results in \HI\ non-detection with a 3-$\sigma$ upper limit presented in Table \ref{SF_LINER_HI_Table}. This result is limited to a small group of objects and needs to be confirmed by larger samples, allowing the detection limit derived for the \HI\ mass to be lowered. 

The lack of optical emission lines along with the \HI\ non-detection and early-type morphology suggest that {\sl inactive} galaxies must represent the true {\sl red and dead galaxy systems} already depleted of cold and ionized gas. Thus, this finding - apart from telling us about the actual mechanism responsible for ionizing the gas in LINERs -  suggests the key importance of gas  (of which \HI\ can represent one of the tracers)  content to make it possible for a galaxy to be classified as LINER. It appears that a galaxy needs to contain gas (cold and/or warm phase) in order to become a LINER, regardless what the mechanism causing the ionized emission is. In addition, the \HI\ properties found for LINERs and inactive galaxies confirm the connection between the presence of neutral and ionized gas, already noted in the study of the \sauron\ sample \citep{Morganti2006}. In the detailed \sauron\ study, a strong link is observed between these two phases of the gas, both in terms of detection rate as well as kinematics. All \sauron\ and LH galaxies where \HI\ is detected also contain ionized gas, whereas no \HI\ is found in galaxies without ionized gas.

\begin{table*}[!]
   \begin{center}
     \begin{tabular}{ l c c c c }
     							  &       &  Radio    &          &  \\
     \hline
    		& Nr ($24 \mu m$), Stacked & SFR$_{1.4 GHz }^{observed}$ [M$_{\odot}$ yr$^{-1}$]    & SFR$_{1.4 GHz}^{24 \mu m }$ [M$_{\odot}$ yr$^{-1}$]    \\
     \hline
         Red LINERs  &  (8), 10      & 1.87 $\pm$  0.09   & 1.74 $\pm$ 0.01    \\
         Blue SF        & (15), 15      &  3.86 $\pm$ 0.15  & 3.58 $\pm$ 0.02  \\
     \\ 		
    \\
      																  &       &  Non-Radio    &          &  \\
   \hline
       		& Nr ($24 \mu m$), Stacked & SFR$_{1.4 GHz }^{observed}$ [M$_{\odot}$ yr$^{-1}$]    & SFR$_{1.4 GHz}^{24 \mu m }$ [M$_{\odot}$ yr$^{-1}$]    \\
   \hline
   Red LINERs       & (6), 14      & -  &   0.40 $\pm$ 0.005           \\
   Blue SF             & (7), 8        & -   &   0.95 $\pm$  0.01        \\
   \\
\end{tabular}
 \caption{Average SFR for red LINER and blue SF galaxies, derived from different SFR indicators in the highest redshift bin (0.06 $<$ $z$ $<$ 0.09). In Col. 2, numbers in the brackets indicate the number of objects for which the 24 $\mu m$ flux is detected above S/N $>$ 2. However, in order to match the total sample contributing to the \HI\ content, the SFRs are averaged over the total number of stacked objects (undetected galaxies are given zero SFR). Errors are calculated from the flux uncertainty of the 24 micron and radio measurements.}
\label{table:SFR}  
\end{center}
\end{table*} 

\begin{figure}[t!]
    \begin{center}
          \includegraphics[width=0.45\textwidth]{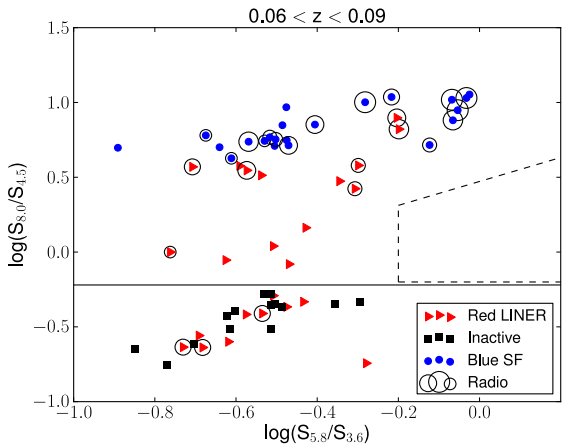}
    \end{center}        
\caption{Infrared color-color plot in the highest redshift bin for galaxies with Spitzer detection (22 blue SF, 24 red LINER and 14 inactive galaxies). Circles are proportional to the radio power distribution. The dashed line indicates the region for powerful AGN.}\label{fig:ColorColor}
\end{figure}		

\begin{figure}[t!]
    \begin{center}
      \includegraphics[width=0.45\textwidth]{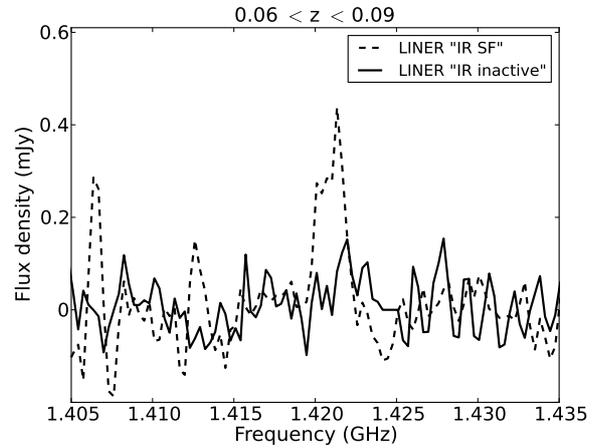}
   \end{center}
   \caption{ \HI\ spectra obtained for LINERs found in the IR SF ({\sl dashed line}) and IR inactive ({\sl solid line}) region. }
\label{fig:Profiles3}
\end{figure}		

\subsection{HI and weak radio sources}\label{Sec:Radio}

As shown in a number of studies, nuclear activity is associated with massive black holes typically hosted by early-type galaxies with massive bulges. In our sample, the most likely group where nuclear activity can be found is the group of red galaxies, as already mentioned in the introduction and discussed in a number of papers (e.g. Kauffmann et al. 2003, and references in Sec. \ref{SampleDescription}).
An active nucleus can reveal its presence in different ways, not only connected with optical emission line properties, and AGN activity can also be indicated by radio emission. However, radio emission can originate from star-formation processes even in early-type galaxies, as it was found for the \sauron\ sample \citep{Oosterloo2010}
Thus, we now turn to investigating, for the group of red galaxies, the nature of the radio emission detected in some of these objects.
Our interest, in particular, is to understand what fraction of the radio emission may be due to nuclear activity. Despite the depth of the radio data available, none of the optically inactive galaxies is found to have a radio counterpart. Thus, we are looking only at the group of red LINERs.

Using the formula by \cite{Yun}, we derive the radio SFR from the observed 1.4 GHz luminosity, and compare it to the {\sl radio SFR inferred from the 24 $\mu$m flux density} (assuming the well known radio/IR correlation holding for SF galaxies, Wu et al. 2005). The 24 $\mu$m continuum is thought to be a good tracer of the warm dust component associated with current star formation in galaxies \citep{Wu}. Under the assumption that all radio emission is due to star-formation processes, the SFR derived from the two indicators (24 $\mu$m and 1.4 GHz ) should be similar, while the presence of a radio AGN would result in a radio excess and a significantly higher value of the observed 1.4 GHz SFR. In Table \ref{table:SFR} we present the SFRs averaged over the total number of stacked objects. The SFRs reveal no significant outliers, suggesting that no significant AGN counterpart is present in our sample. In the same table, for a comparison the values for blue SF galaxies are also indicated. 

It is also interesting to compare the star-formation properties of the radio and non-radio samples. In Table 2, the SFR of radio-detected sources (both LINERs and SF galaxies) is always higher compared to non-radio galaxies.

The results above indicate that the radio emission in these sub-mJy radio sources at relatively low redshift (up to $z=0.09$) is arising mostly from SF. As in early-type galaxies in the nearby \sauron\ sample \citep{Oosterloo2010}, the radio detection of LH galaxies is mostly connected to the level of star formation i.e. more enhanced SFR in galaxies with radio counterparts. In the \sauron\ sample, galaxies with \HI\ in the central regions were found to be more likely detected in radio continuum. This was shown to be due to a higher probability for star formation to occur in galaxies with \HI\ gas, and not to \HI-related AGN fuelling. This could be the case also in the LH sample and we explore this more in the next session.

To further confirm the absence of any powerful AGN, we have created the infrared color-color plot based on Spitzer colors, first proposed by \cite{Lacy} and then revisited by \cite{Stern}. The resulting plot is shown in Fig. \ref{fig:ColorColor}, illustrating that the red LINERs studied here - and more in particular the radio detected -  appear to avoid the region of powerful AGN, indicated with a dashed line. Although the color-color plot confirms that no powerful AGN are present, several studies based on X-ray selected samples suggest that low (X-ray) luminosity or obscured AGN are missed by the IR color-color selection wedge, and tend to appear in the region of inactive galaxies \citep{Cardamone, Brusa, Eckart}. A similar result was obtained by \cite{Prandoni2009}, who found that radio-selected AGN, typically hosted by elliptical galaxies, can be located both in the AGN region (defined by the dashed line in Fig. \ref{fig:ColorColor}) and in the region of inactive galaxies. Thus, based on this plot, we can not yet completely exclude the presence of weak AGN in, at least, a subset of the red galaxies in the LH sample. 

\subsection{\HI\ and 24 $\mu$m emission in LINERs: are there two groups in terms of SF?}\label{Sec:TwoGroupLINER}

We investigate now in some more detail the apparent split of the red galaxies (red LINERs and inactive) around the line corresponding to $log(S_{8.0}/S_{4.5}) = -0.2$ in Fig. \ref{fig:ColorColor}. This rather arbitrary separation does actually correspond to the 24 $\mu$m detection of the sources, i.e above the line all but two (one blue SF and one red LINER) sources are detected in 24 $\mu$m, whereas below the line all but one (one red LINER) galaxies have no associated 24 $\mu$m dust emission. 

Interesting is the connection between the detection/non-detection of the 24 $\mu$m star formation tracer and the \HI\ detections. The stacked profiles and \HI\ content of the two groups of red (LINER) galaxies are presented in Fig. \ref{fig:Profiles3} and Table \ref{SF_LINER_HI_Table}, the two regions being referred to as {\sl IR SF} region and {\sl IR inactive} region. LINERs in the IR SF region are detected with relatively high \HI\ content and often have associated radio counterparts. However, the LINER group in the IR inactive region does not show any \HI\ detection and is largely populated by non-radio LINERs.

These properties indicate a strong correlation between the presence of 24 $\mu$m emission and \HI\ content, pointing towards SF-driven 1.4 GHz emission in the IR SF region, as seen in Sec. \ref{Sec:Radio}. This is consistent with what was found for early-type galaxies in the nearby Universe (see the study of the \atlas\ sample by Serra et al. 2012) where the neutral hydrogen seems to provide material for star formation, and galaxies containing central \HI\ exhibit signatures of ongoing star formation five times more frequently than galaxies without central \HI. For the group of radio LINERs in the IR inactive region, the lack of \HI\ and 24 $\mu$m dust emission suggests that radio emission is less likely arising from SF, but could be, indeed, related to weak AGN activity.

These LH results suggest that, using IR information, LINERs can be disentangled into two groups. One group is actively star-forming, while other LINERs show more resemblance to optically inactive galaxies and are the best candidates for hosting low-luminosity AGN.

\section{Summary and Conclusions}

This study has investigated the \HI\ content - using stacking techniques - of the galaxies in the LH area, observed with the WSRT telescope and combined with SDSS spectra. We have focused our analysis on the redshift bin 0.06 $\textless$ $z$ $\textless$ 0.09, i.e. a redshift range relatively clean from selection biases.  
The main results of this study can be summarized in: \\
-- Both red and blue galaxies are detected in the \HI\ stacked profile. Blue galaxies are more \HI\ rich, but also red galaxies show interesting amounts of \HI. \\
-- Inactive (in terms of optical emission lines) galaxies are not detected in \HI. This group of early-type galaxies appears to be genuinely depleted of cold and ionized gas.\\
-- The \HI\ properties found for red LINERs and inactive galaxies confirms the strong connection between the presence of neutral and ionized gas, already noted in the study of the \sauron\ sample \citep{Morganti2006}. All galaxies where \HI\ is detected also contain ionized gas, whereas no \HI\ is found around galaxies without ionized gas. \\
-- For the majority of {\sl radio} LINERs, the radio emission appears to be connected with enhanced star formation.\\
-- LINERs can be separated into two groups based on 24 $\mu$m emission properties. LINERs detected at 24 $\mu$m show relatively large amounts of \HI\ and are often detected in radio as a result of ongoing SF. The lack of \HI\ for the group of 24 $\mu$m  undetected LINERs points towards resemblance with optically inactive galaxies. Radio LINERs in the latter group are the best candidates for hosting low luminosity AGN.  
\\
\\ 
This study has been limited by the number of objects and by the available SDSS redshifts. The next step will be to increase by at least an order of magnitude the available objects. To do this, an observational campaign is in progress to extend the area covered by observing many fields. With a larger number of sources, the noise level in the stacked spectra can be significantly reduced, making it possible to study various samples, and to detect \HI\ down to lower limits compared to this study. Furthermore, we expect that in a larger sample the number of potential AGN (optical and/or radio) will increase, making it possible to further investigate the connection between the central activity and gas.

\section{Acknowledgements}
We thank the reviewer for the useful and detailed comments that helped us to improve the manuscript. 

Funding for the SDSS and SDSS-II has been provided by the Alfred P. Sloan Foundation, the Participating Institutions, the National Science Foundation, the U.S. Department of Energy, the National Aeronautics and Space Administration, the Japanese Monbukagakusho, the Max Planck Society, and the Higher Education Funding Council for England. The SDSS Web Site is http://www.sdss.org/.

The SDSS is managed by the Astrophysical Research Consortium for the Participating Institutions. The Participating Institutions are the American Museum of Natural History, Astrophysical Institute Potsdam, University of Basel, University of Cambridge, Case Western Reserve University, University of Chicago, Drexel University, Fermilab, the Institute for Advanced Study, the Japan Participation Group, Johns Hopkins University, the Joint Institute for Nuclear Astrophysics, the Kavli Institute for Particle Astrophysics and Cosmology, the Korean Scientist Group, the Chinese Academy of Sciences (LAMOST), Los Alamos National Laboratory, the Max-Planck-Institute for Astronomy (MPIA), the Max-Planck-Institute for Astrophysics (MPA), New Mexico State University, Ohio State University, University of Pittsburgh, University of Portsmouth, Princeton University, the United States Naval Observatory, and the University of Washington.

This work is based [in part] on observations made with the Spitzer Space Telescope, which is operated by the Jet Propulsion Laboratory, California Institute of Technology under a contract with NASA.

\end{document}